\documentclass[twocolumn]{aastex62}

\graphicspath{{./}{figures/}}

\received{}
\revised{}
\accepted{}

\submitjournal{AJ}

\shorttitle{Analysis of membership probability in nearby young moving groups with $\textit{Gaia}$ DR2}
\shortauthors{Ujjwal et al.}
\usepackage{subfigure}
\usepackage{mathtools}
\usepackage{multirow}\usepackage[normalem]{ulem}
\useunder{\uline}{\ul}{}

\begin{document}

\title{Analysis of membership probability in nearby young moving groups with $\textit{Gaia}$ DR2}

\correspondingauthor{Sreeja S. Kartha}
\email{sreeja.kartha@christuniversity.in}

\author{K. Ujjwal}
\affil{Department of Physics and Electronics, CHRIST (Deemed to be University), Bangalore 560029, India}
\author{Sreeja S. Kartha}
\affil{Department of Physics and Electronics, CHRIST (Deemed to be University), Bangalore 560029, India}
\author{Blesson Mathew}
\affil{Department of Physics and Electronics, CHRIST (Deemed to be University), Bangalore 560029, India}
\author{P. Manoj}
\affiliation{Department of Astronomy and Astrophysics, Tata Institute  of Fundamental Research, Homi Bhabha Road, Colaba, Mumbai 400005, India}
\author{Mayank Narang}
\affiliation{Department of Astronomy and Astrophysics, Tata Institute of Fundamental Research, Homi Bhabha Road, Colaba, Mumbai 400005, India}

\begin{abstract}

We analyze the membership probability of young stars belonging to nearby moving groups with $\textit{Gaia}$ DR2 data. The sample of  1429 stars were identified from `The Catalog of Suspected Nearby Young Moving Group Stars'. Good-quality parallax and proper motion values were retrieved for 890 stars from $\textit{Gaia}$ DR2 database. The analysis for membership probability is performed in the framework of LACEwING algorithm. From the analysis it is confirmed that 279 stars do not belong to any of the known moving groups. We estimated the {\it U, V, W} space velocity values for 250 moving group members, which were found to be more accurate than previous values listed in the literature. The velocity ellipses of all the moving groups are well constrained within the ``good box", a widely used criterion to identify moving group members. The age of moving group members are uniformly estimated from the analysis of {\it Gaia} Color-Magnitude Diagram with MIST isochrones. We found a spread in the age distribution of stars belonging to some moving groups, which needs to be understood from further studies.

\end{abstract}

\keywords{stars: moving group -- proper motion --space velocity   }

\section{Introduction}           
\label{sect:intro}

Moving groups are sparse and loosely bound groups of stars, which are coeval and spread over thousands of cubic parsecs in space \citep{Eggen1996}. The origin of moving groups still remains as an interesting puzzle. Present understanding about the formation of moving groups proposes three possible mechanisms, i.e., i) cluster disruption, ii) dynamical perturbations, and iii) satellite accretion. Cluster disruption considers moving groups as a dispersed cluster. One of the likely scenario for the disruption can be an encounter of an open cluster with giant molecular clouds or massive objects in the Galaxy \citep[e.g.,][]{Asiain1999}.  The theory of dynamical perturbations suggests that some of the moving groups are formed as a result of the dynamical perturbations introduced by the non-axisymmetric components of the galaxy such as bar/spiral arms \citep[e.g.,][]{Antoja2010}. According to the satellite accretion theory, the Milky Way accretes its dwarf satellite galaxies and the debris of such accreted satellites remain in the Galaxy as stellar streams or moving groups. Although there are discussions which disfavour the compatibility of the satellite accretion theory, \citep[e.g.,][]{Famaey2008}, it is still considered as one of the possible mechanisms for the formation of moving groups \citep[refer][]{Antoja2010, Ramya2012}. The study of stars in nearby young moving groups (NYMG) thought to originate from clusters has attracted lots of attention since they are prime laboratories to understand the star formation and cluster dispersion mechanisms \citep{Zuckerman2004,Torres2003,Torres2008}. There were prominent surveys/programs designed to study various aspects of moving groups, as for example, `Search for Association Containing Young stars' \citep[SACY;][]{Torres2006, Viana2009, dasilva2009, desilva2013, Elliot2014, Elliot2015, Elliot2016} and `All-sky Co-moving Recovery Of Nearby Young Members' \citep[ACRONYM;][]{Kraus2014, Shkolnik2017, Schneider2019A}. 

Young moving group members are dynamically similar to their parental distribution. Therefore, the members of a moving group are identified from their common space velocities. \citet{Mamajek2005} suggested `convergent point analysis' to identify the region of formation of members belonging to a moving group. This method is also useful to identify new members which share similar space motion with a known moving group \citep[e.g.][]{Torres2006,Rodriguez2011}. Recent researches also makes use of this method \citep[e.g.][]{Roser2019}. The studies by \cite { Meingast2019_1} and \cite{ Meingast2019} make use of a new method of identifying cluster members by making use of  3D galactocentric cylindrical velocity space. Various algorithms were developed to estimate the membership probability of a star belonging to a moving group. Some of the prominent algorithms/codes widely used in the literature are BANYAN (and the advanced versions), LACEwING and Chronostar \citep{Chronostar2019}. `Bayesian Analysis for Nearby Young AssociatioNs' (BANYAN) is a Bayesian inference based algorithm which includes the kinematic information of nearest ($<$100 pc) and youngest ($\sim$ 10$-$200 Myr) moving groups \citep[e.g.][]{Malo2013,Gagne2014,BANSIG2018}. `LocAting Constituent mEmbers In Nearby Groups' (LACEwING) is a moving group analysis code, built on the kinematic information of 16 associations and open clusters in the age range $\sim$5--800 Myr \citep{Riedel2017}. With the advent of space missions such as {\it Gaia}, accurate and reliable astrometric and photometric data is becoming available, which can be used to confirm the membership of stars in nearby young moving groups. 

\cite{Gagne2018} and \cite{Faherty2018} studied the nearby young stars and their possible association with moving groups using the Tycho-$\textit{Gaia}$ Astrometric Solution (TGAS). For this work, they used data resources from $\textit{Gaia}$ Data Release 1 (DR1), which lacked photometry and hence needed to be complimented from different surveys. These studies highlighted the capability of \textit{Gaia} mission and proposed that \textit{Gaia} DR2 will provide a more clear idea about the membership probability for every star. One of the initial studies based on data from $\textit{Gaia}$ DR2 was performed by \cite{GagneDR22018ApJ...862..138G}, who selected all the \textit{Gaia} DR2 sources within 100 pc and using BANYAN $\Sigma$ tool identified 898 new highly likelihood members. 

The objective of the present study is to use the best available astrometric and photometric data provided by the $\textit{Gaia}$ DR2 to estimate the membership and properties of moving groups. For this work we employed all the thirteen NYMGs identified till date, namely, $\epsilon$ Chamaeleontis \citep[$\epsilon$ Cha;][]{Mamajek2000, Murphy2013}, Hercules-Lyra \citep[Her-lyr;][]{Gaidos1998, Fuhrmann2004}, TW Hydra \citep[TW Hya;][]{delaReza1989}, $\beta$ Pictoris \citep[$\beta$ Pic;][]{Barrado1999}, Octans \citep{Torres2008,Murphy2015}, Tucana - Horologium \citep[Tuc-Hor;][]{Torres2000,zuckerman2001}, Columba \citep{Torres2008}, Argus \citep{Torres2008,desilva2013}, AB Doradus \citep[AB Dor;][]{Torres2003, Bell2015}, Ursa Major \citep{King2003}, $\chi^{01}$ For \citep{Dias2002}, Carina \citep{Torres2008} and Carina-near \citep{Zuckerman2006}. The sample of stars used for this study is explained in Sect. 2. In Sect. 3, we have discussed the need for the re-analysis of membership probability of NYMGs in the context of the availability of {\it Gaia} DR2 data. Also, we have estimated the {\it U, V, W} space velocities and ages of the stars in the moving groups. The major results from this study are summarized in Sect. 4. 

\section{Data Analysis with $\textit{Gaia}$ DR2} 

$\textit{Gaia}$ is an ongoing legacy mission of European Space Agency (ESA), designed to provide accurate distance, proper motion and radial velocity measurements of more than a billion stars in our Galaxy. $\textit{Gaia}$ was successfully launched on 19 December 2013.  After completing the commissioning period in the summer of 2014, the five-year nominal science phase operation started. Two sets of catalogs have been produced by $\textit{Gaia}$ till date.  $\textit{Gaia}$ Data Release 2 ($\textit{Gaia}$ DR2) was released on 25 April 2018, based on 22 months of observations made between 25 July 2014 to 23 May 2016. $\textit{Gaia}$ DR2 includes the position, parallax, and proper motion measurements \citep{dr2018}, red and blue photometric data for about 1.3 billion stars \citep{Evans2018} and radial velocity values of about 7 million stars \citep{Cropper2018}.

The `Catalog of Suspected Nearby Young Moving Group Stars' (CSNYMGS) intends to serve as a comprehensive collection of stellar parameters of almost all nearby young astrophysical objects, including, stars, planetary-mass objects and brown dwarfs, ever reported as young \citep{Riedel2016,Riedel2017}. The catalog \footnote{\flushleft{$ github.com/ariedel/young_catalog/blob/master/ $-$ young.master.2017.0601.csv$}} contains photometric and spectroscopic information of 5401 objects from surveys/missions such as Hipparcos \citep{hipp2007}, $\textit{Gaia}$ DR1 \citep{Gaia2016} and `The American Association of Variable Star Observers Photometric All Sky Survey' \citep[APASS DR9;][]{Henden2016}. For this study we include a sample of 1429 stars from CSNYMGS, which are identified as members of the known moving groups. By making use of the 2MASS IDs given in the catalog, the $\textit{Gaia}$ DR2 data are extracted \citep[see][]{Marrese2019}. Sources without parallax were removed from the cross-matched data. After the cross-match and removing the sources without parallax, $\textit{Gaia}$ DR2 data for 1205 stars were obtained. It is to be noted that 224 stars which were found to be missing from the total sample, which were either non-detections or does not have parallax measurements. Most of these stars were identified as brown dwarfs or high proper motion stars in literature. Re-normalised Unit Weight Error (RUWE, a parameter estimated using the chi squared of the astrometric fit and stellar colour) is considered for confirming the quality of the {\it Gaia} DR2 data \citep{Lindegren2018}. Only stars with RUWE $<$ 1.4 is used for the study. After this iteration, we got {\it Gaia} DR2 astrometric and photometric data for 890 stars in which 367 stars have {\it Gaia} DR2 radial velocity measurements. These 890 stars belonging to thirteen moving groups will be used for the present work.

\section{Results and Discussion}

\subsection{Analysis of moving group membership with LACEwING using \textit{Gaia} DR2 data}

LACEwING uses the spatial and kinematic information of an object such as Right Ascension ($\alpha$), Declination ($\delta$), distance, proper motion in Right Ascension ($\mu_{\alpha}$), proper motion in Declination ($\mu_{\delta}$), and radial velocity ($\gamma$) to check whether the object is a member of any of the known NYMGs. Inside the LACEwING framework each NYMG is considered as two triaxial ellipsoids having base values of {\it U, V, W} and {\it X, Y, Z}. For the members of each moving group, LACEwING predicts the observable values such as distance, $\mu_{\alpha}$ and $\mu_{\delta}$ at the $\alpha$ and $\delta$ of the input star. These predicted values are compared with the measured quantities. The working of LACEwING mainly comprises of three steps. In the first step, LACEwING predicts the properties such as proper motion, radial velocity and distances corresponding to  $\alpha$ and $\delta$. To generate the spread of {\it U, V, W} space velocities of each NYMG, 100,000 Monte-Carlo iterations are made within the triaxial ellipsoids to find out the actual membership probability. The {\it U, V, W} velocities are transformed into proper motion components, $\mu_{\alpha}$, $\mu_{\delta}$ and $\gamma$ by inverted matrices given in \cite{Johnson1987} to simulate a star for each moving group at a standard distance of 10 parsecs for the same $\alpha$ and $\delta$ of the target. By making use of this predicted values and the available kinematic data of the star, four goodness-of-fit matrices such as proper motion, distance, radial velocity and position will be produced. The number of matrices can vary upto four based on the availability of observables of input star. The quality of match of the star with one of the moving group is represented by each of these matrices. In the second step, these matrices are combined in quadrature and the goodness-of-fit values are converted into membership probabilities. Also LACEwING accounts all the 7 possible combinations of $\alpha,  \delta$, $\mu_{\alpha}$, distance, $\mu_{\delta}$ and $\gamma$ each having its own goodness-of-fit values. These probabilities are quantified by making use of a Solar Neighborhood simulation which is ran through the LACEwING algorithm. Hence, even if one of the parameter is missing, the other combinations will be used to estimate the membership probability.

Further analysis of membership probability is carried out on this simulated star \citep{Riedel2017}.  A small variation in $\mu_{\alpha}$, $\mu_{\delta}$ can reflect in the distance, {\it U, V, W} and {\it X, Y, Z} values. This can easily be evaluated by comparing new estimations from the \textit{Gaia} DR2 data with those listed in CSNYMGS. 

\begin{figure}[h]
\begin{center}
\includegraphics[width = 1\columnwidth]{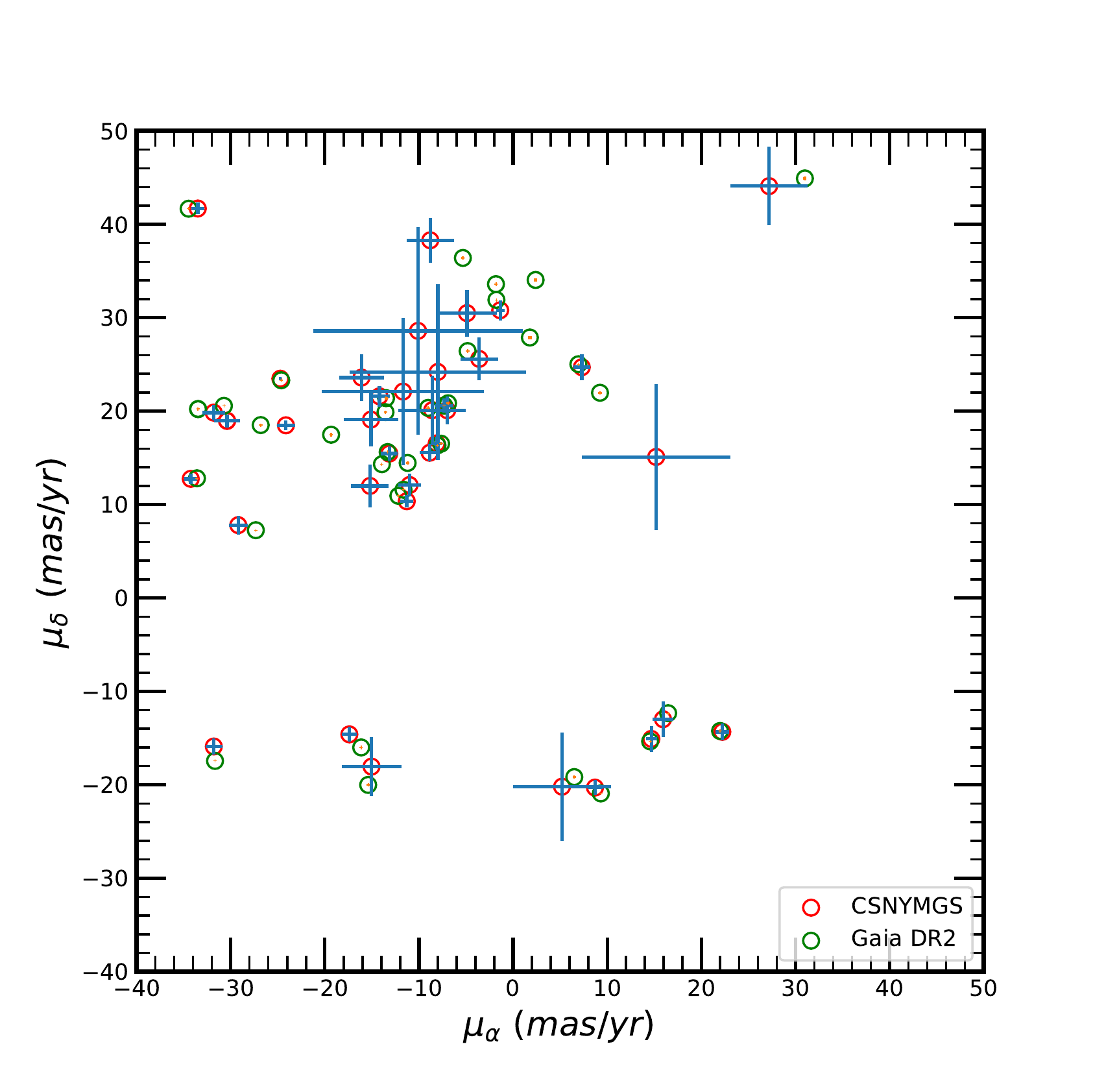}
\caption{Vector point diagram depicting the proper motion of stars in Octans moving group: Red circle and blue cross represents the proper motion and the associated error from CSNYMGS \citep{Riedel2017}, respectively. Green circles and orange cross represents the proper motion and the associated error from the $\textit{Gaia}$ DR2, respectively. Accuracy of $\textit{Gaia}$ DR2 data over previous estimates is evident from the figure.}
\label{fig:vector}
\end{center}
\end{figure}

Vector point diagram is one of the basic tools that can be used to distinguish moving group members from field stars \citep[e.g.,][]{Hoogerwerf1999}. Figure \ref{fig:vector} represents the vector point diagram of stars in Octans moving group, constructed using proper motion information from $\textit{Gaia}$ DR2. Also shown are the proper motion values of the members listed in CSNYMGS. It is evident from the figure that the proper motion values from the $\textit{Gaia}$ DR2 shows less error when compared to the values listed in CSNYMGS. This explains the accuracy of {\it Gaia} DR2 proper motion values over previous estimates.

Since our moving group stars are close to the Sun, they are spread over a large area in the sky. Hence, it is difficult to identify their association by means of stellar density, as in the case of star clusters. The best way to group these stars is based on their common space motion \citep{Torres2006}. The inclusion and elimination of members from a moving group rely on the space velocity values, for which we need accurate parallax, proper motion and radial velocity measurements \citep{Zuckerman2004}. $\textit{Gaia}$ DR2  currently  provides the best possible proper motion and radial velocity values, from which we can calculate the space velocity ({\it U, V, W}) values. We selected $\mu_{\alpha}$, $\mu_{\delta}$ and $\gamma$ from $\textit{Gaia}$ DR2 database for our sample of 890 stars. Since 367 stars were having {\it Gaia} radial velocity measurement, we have used this sub-sample for this particular analysis. The {\it U, V, W} space velocities are calculated using the matrices from \cite{Johnson1987}, in which right-handed convention is used. This means that {\it U, V, W} values are considered positive in the direction of Galactic center, Galactic rotation and the North Galactic Pole (NGP), respectively. Along with the Right Ascension ($\alpha$) and Declination ($\delta$) of the objects, the matrices make use of the following quantities and their uncertainties, i.e., the parallax  in arcsec ($\pi\pm\sigma_{\pi}$),
the radial velocity in $km~s^{-1}$ ($\gamma\pm\sigma_{\gamma}$),
the proper motion in right ascension in $arcsec~yr^{-1}$ ($\mu_{\alpha}\pm\sigma_{\mu_{\alpha}}$) and the proper motion in declination in $arcsec~yr^{-1}$ ($\mu_{\delta}\pm\sigma_{\mu_{\delta}}$). The matrices, as given in \cite{Johnson1987}, are reproduced below.

The transformation matrix is given by,\\
\[
\bf{T} = 
\begin{bmatrix}
-0.06699 & -0.87276 & -0.48354 \\
+0.49273 & -0.45035 & +0.74458 \\
-0.86760 & -0.18837 & +0.46020
\end{bmatrix}
\]
and the coordinate matrix is defined as,
\[ 
\bf{A} =
\begin{bmatrix}
+\cos\alpha \cos\delta & -\sin\alpha & -\cos\alpha\sin\delta \\
+\sin\alpha\cos\delta & +\cos\alpha & -\sin\alpha\sin\delta \\
+\sin\delta & 0 & +\cos\delta
\end{bmatrix}
\]
The space velocity components are given by,
\[
\begin{bmatrix}
U \\
V \\
W
\end{bmatrix}
=
\bf{B} 
\begin{bmatrix}
\gamma \\
\frac{k\mu_{\alpha}}{\pi}\\
\frac{k\mu_{\delta}}{\pi}
\end{bmatrix}
\]
where {\bf B} = {\bf T}$\cdot${\bf A}, and k = 4.74057, the equivalent in km $s^{-1}$ of one astronomical unit in one tropical year.
The associated error is estimated by,

\[
\begin{bmatrix}
\sigma_{U}^{2} \\
\sigma_{V}^{2} \\
\sigma_{W}^{2}
\end{bmatrix}
=
\bf{C}
\begin{bmatrix}
\sigma_{\gamma}^{2} \\
(k/\pi)^{2} [\sigma_{\mu_{\alpha}}^{2}(\mu_\alpha\sigma_{\pi}/\pi)^{2} \\
(k/\pi)^{2} [\sigma_{\mu_{\delta}}^{2}(\mu_\delta\sigma_{\pi}/\pi)^{2}
\end{bmatrix}
\]
\[+
2\mu_{\alpha}\mu_{\delta}k^{2}\sigma_{\pi}^{2}/\pi^{4}
\begin{bmatrix}
b_{12}\cdot b_{13}\\
b_{22}\cdot b_{23}\\
b_{32}\cdot b_{33}
\end{bmatrix}
\]
where, $c_{ij} = b_{ij}^{2}$ for all i and j.

As an example, the {\it U, V, W} values calculated for stars in the ABDor moving group are given in Table 1. Also shown are the values listed in CSNYMGS.  The average percentage error associated with the previous {\it U, V, W} estimates are 14$\%$, 4.8$\%$ and 17.2$\%$, respectively. But the average percentage error obtained from our estimate using {\it Gaia} DR2 reduced to 6.4$\%$, 1.9$\%$ and 5.1$\%$, respectively. Since the error associated with the estimates reduced significantly it is clear that the {\it U, V, W} space velocities of stars estimated from {\it Gaia} DR2 are accurate than previous estimates.

\begin{figure}[h]
\begin{center}
\includegraphics[width = 1\columnwidth]{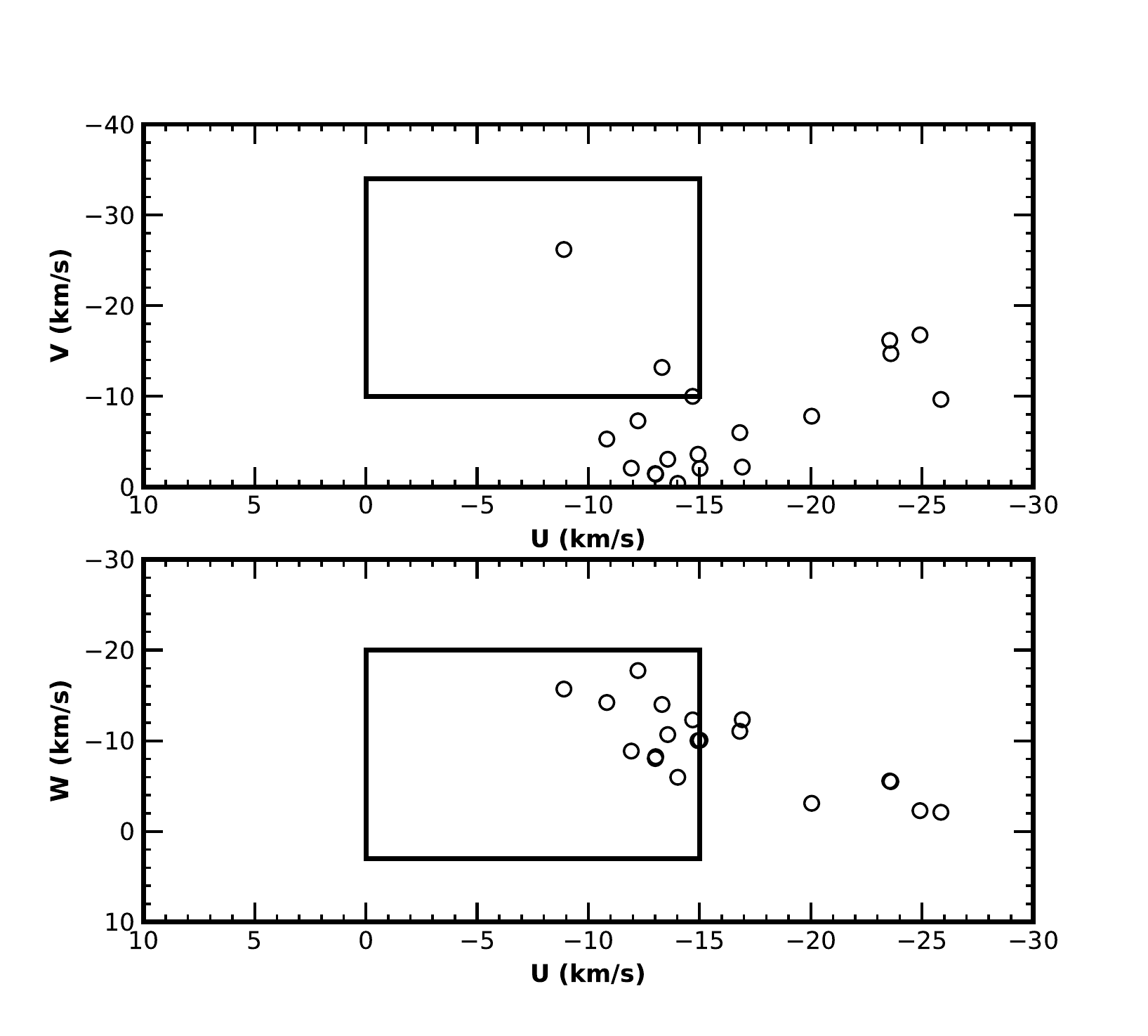}
\caption{Space velocity distribution of stars in Octans moving group calculated using the {\it Gaia} DR2 data. Boxes represents the region specified by the ``good box" criterion. It is seen that the ``good box" criterion is not completely satisfied by the known members in the Octans moving group.}
\label{fig:SpaceVel}
\end{center}
\end{figure}

Based on the kinematic analysis of moving group members, \cite{Zuckerman2004} suggested that {\it U, V, W} velocities of almost all the young stars are within 10 km~s$^{-1}$ of the average {\it U, V, W} value of Eggen's Local Association \citep{Jeffries1995}. It is then generalized that, for young stars {\it U, V, W} velocities fall in the ``good box" with {\it U, V, W} dimensions in the range 0 -- -15 km~s$^{-1}$, -10 -- -34 km~s$^{-1}$, and +3 -- -20 km~s$^{-1}$, respectively \citep{Zuckerman2004}. We checked whether our sample of moving group candidates fall inside the ``good box" while using {\it Gaia} DR2 data. As an example, we have defined the ``good box" for the moving group Octans in Figure \ref{fig:SpaceVel}. It is evident from the figure that the {\it U, V, W} velocities of the moving group members calculated from the {\it Gaia} DR2 data spans out of the ``good box". Hence, the ``good box" criterion is not completely satisfied by the catalogued members in the Octans moving group. This suggests the need to re-estimate the membership of moving group candidates using {\it Gaia} DR2 data.  

\subsection{Estimation of membership probability using $\textit{Gaia}$ DR2 data}

We have analyzed the membership probability of candidates in the moving groups with LACEwING algorithm using $\textit{Gaia}$ DR2 data. LACEwING is used to calculate the membership probability of a star in a particular moving group \citep{Riedel2017}. Since LACEwING is capable of determining the membership probability even with the incomplete kinematic data, by making use of the goodness-of-metrices and the simulation of solar neighbourhood as discussed in section 3.1, we were able to reanalyse the membership probability of the 890 stars. During the analysis we found the membership of some stars changed from one moving group to another. As an example, we explain the case of Tuc-Hor moving group, which hosts maximum number of stars. Although Horologium moving group \citep{Torres2000} and Tucana moving group \citep{zuckermanetal.2001} were discovered separately, \cite{zuckerman2001} confirmed that both of them contain similar stars and hence was classified as Tucana-Horologium (Tuc-Hor) moving group. Our analysis suggested that most of the members in Tuc-Hor has a definite probability to belong to either Columba or $\beta$ pic moving groups. This may be due to the overlap in the properties of these moving groups. One important aspect we saw during the analysis is that a few stars which were identified as moving group members now belong to open clusters. The stars RX J1005.3-7749 and RX J0915.5-7609 are now confirmed as $\eta Cha$ members whereas 2MASS J08082487+4347557, HD 221539, 2MASS J02501167-0151295, HIP 15247A, HIP 15247B, HD 22213, 2MASS J03445673-1145126, 2MASS J04174743-2129191 and 2MASS J04365738-1613065 are re-assigned to Hyades open cluster. This can be seen as a direct consequence due to the change in the input data used for the present analysis.

From the analysis of membership probability using {\it Gaia} DR2 data, we found 611 stars remaining as members of any of the thirteen moving groups. It may be noted that from \citet{Riedel2017} we have obtained 890 candidates as moving group members having {\it Gaia} DR2 data. Hence, from our analysis by making use of accurate {\it Gaia} DR2 data, we found that 279 stars do not belong to any of the known moving groups. The new estimates of distance, age and space velocity values of moving group members from the available {\it Gaia} DR2 is listed in Table 2. So, for the present study, we will be using a sample of 611 member stars, belonging to 13 moving groups.

\subsection{Space velocity analysis of the moving group members}

The analysis of space velocity plays a key role in the membership determination of stars in a moving group. This is further important in deciding the ``good box" criterion for moving groups \citep{Zuckerman2004}. $\textit{Gaia}$ DR2 provides homogeneity in the measurement of parameters such as parallax, proper motion and radial velocity, which is one of the fundamental requirement for estimating $U$, $V$, $W$ space velocity values. We found that among the moving group candidates listed in CSNYMGS, 250 have parallax, proper motion and radial velocity values in the {\it Gaia} DR2, which can be used for estimating space velocity. It may be noted that the number of stars for this analysis got reduced since only 250 stars have radial velocity values from {\it Gaia} DR2. 

We estimated the {\it U, V, W} values of 250 stars using the method described in previous section (Sec. 3.1). The {\it U, V, W} estimates from this study shows considerably lesser error with a mean value of around 0.54 km~s$^{-1}$, 0.76 km~s$^{-1}$ and 0.58 km~s$^{-1}$, respectively, which is better than previous reported values \citep{montes2001,Malo2013,Maldonado2010,Shkolnik2012,Makarov2007}. For example, the average error estimate in the space velocity values for the re-analyzed members of AB Dor moving group is around 0.32 km~s$^{-1}$ whereas those compiled in CSNYMGS has an average error of 1 km~s$^{-1}$. This is because the input values for space velocity estimation were taken from different resources in previous studies. For example, \cite{Malo2013} compiled parallax, proper motion and radial velocity values from various resources such as \cite{hipp2007}, \cite{Koen2010}, \cite{Messina} and \cite{Bailey2012}. The advantage of present space velocity estimation is due to the homogeneous dataset of {\it Gaia} DR2, from which all the input data is taken. In addition, the input values such as proper motion and radial velocity are more accurate than previous missions/surveys. 

\begin{figure*}[h]
\begin{center}
\includegraphics[width = 1\textwidth]{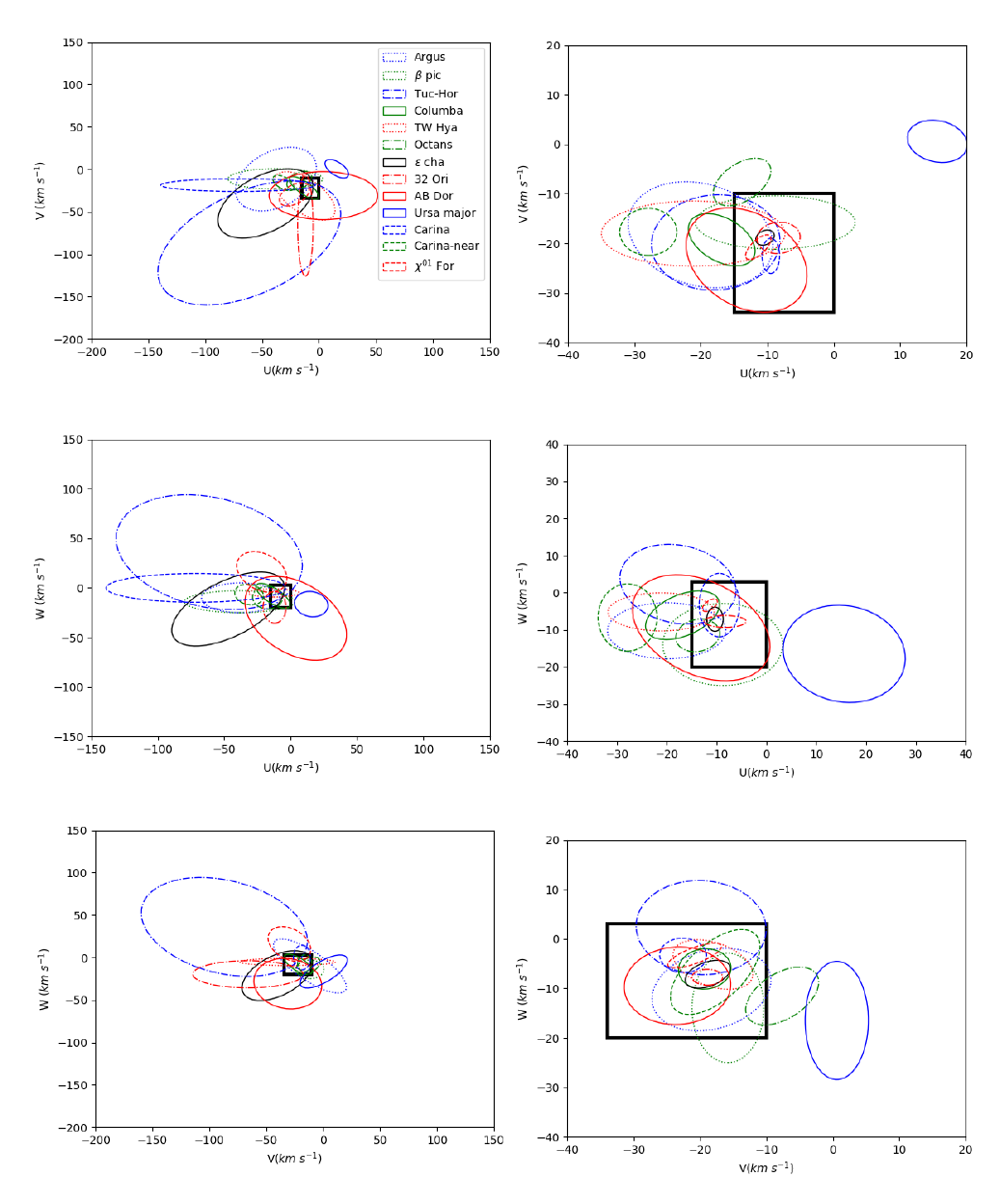}
\caption{Representation of moving group members in the space velocity plane. Left panel represents the initial sample of 367 stars and right panel denotes the 250 stars obtained after the re-analysis with {\it Gaia} DR2. Each moving group is color coded, which is presented in the label of top left figure. The spread is indicated by the extend of the ellipse. The ``good box" is also represented in the figure with solid black line.}
\label{fig:VelEllipse}
\end{center}
\end{figure*}

The velocity ellipses between the different space velocity components ($U$ and $V$, $U$ and $W$, and $V$ and $W$) are generated for all the moving groups and is shown in Figure \ref{fig:VelEllipse}. The left panel shows the ellipse generated using a sample of 367 stars from CSNYMGS having $U, V, W$ velocity estimates from {\it Gaia} DR2. The right panel represents the space velocities of the new sample of 250 members whose membership is confirmed from this study. We have included the ``good box" criterion in the figure with the dimensions 0 -- -15 km~s$^{-1}$, -10 -- -34 km~s$^{-1}$, and +3 -- -20 km~s$^{-1}$ for $U$, $V$ and $W$ respectively. It is seen from Figure \ref{fig:VelEllipse} that the velocity ellipses of all the moving groups are well constrained within the ``good box". Hence, from the analysis of space velocities, we found that 117 stars (i.e., 367 - 250) do not belong to any of the known moving groups. We had shown in previous section that 279 stars are identified as non-members from the analysis of proper motion in LACEwING framework. Among this sample, the space velocity analysis has reconfirmed the non-membership of 117 stars.
In the {\it Gaia} era, there is a need to re-estimate the ``good box" criterion. Since, LACEwING makes use of ``good box" defined in \cite{Zuckerman2004} for estimating the membership probability, we are also considering the same criterion for the present work. Distribution of space velocity estimates obtained for the confirmed moving group stars suggests a more wider ``good box" compared to that proposed by \cite{Zuckerman2004}. However, as a future work, we plan to re-define the ``good box" criteria by making use of {\it Gaia} DR 3 data.

\subsection{Age estimation}

Stars belonging to nearby young moving groups are quite near ($<$ 100 pc) and hence the protoplanetary disc in them can be studied in detail using imaging techniques such as interferometry \citep [e.g.,][]{Rebull2008, Zuckerman2011, Rodriguez2015, Binks2017}. The proximity of these stars makes them the best targets for direct detection of low-mass planetary companions \citep [e.g.,][]{Chauvin2004,Chauvin2005,Lagrange2010, Bowler2013,Bowler2017, Dupuy2018}. For studies related to protoplanetary disc and planetary mass companions, it is important to accurately estimate the age of the host star. 

Conventionally, the age of a star is estimated from their location in the Color-Magnitude Diagram (CMD), which is overplotted with isochrones generated from stellar models \citep{ortega2002,ortega2004, Song2003}. Age can also be estimated by other means such as Lithium abundance, rotation rate and stellar activity \citep{Zuckerman2004,dasilva2009}. Age estimates based on Lithium Depletion Boundary (LDB) technique was proposed as one of the most reliable age estimation methods since it is less model dependent compared to other age estimation techniques \citep{Soderblom2014, Mathew2017}. However, LDB technique provide reliable age estimates for stars in which lithium burning happens in the early phase of their lifetime. This restricts the spectral range over which this technique can be used. The dynamical age of a moving group can also be estimated from convergent point analysis \citep[e.g.][]{Barenfeld2013}. \citet{Bell2015} employed semi-empirical pre-main sequence models with $\tau^{2}$ minimization to estimate the age of moving groups. Most of these studies estimate the common age of the moving group, considering the fact that all stars which belong to moving group are {\it bona fide} members. It is to be noted that there are issues related to the model dependency of age estimates using isochrone fitting \citep{Bell2012, Bell2013, Bell2014} when factors of stellar variability and binarity are taken into account. But the high quality photometry and distance data available from $\textit{Gaia}$ DR2 enable us to create better CMD. We have estimated the individual age of the moving group members from {\it Gaia} CMD.


The science data of $\textit{Gaia}$ DR2 consists of broad-band photometry in the unfiltered G band and integrated $G_{BP}$ and $G_{RP}$ magnitudes \citep{Gaia2016}. This will equip us with a complete data set for constructing CMD purely based on $\textit{Gaia}$ DR2 data \citep{GaiaHR2018}. It may be noted that the advantage of this CMD analysis is due to the homogeneous use of stellar parameters from a single mission/database. By making use of distances estimated from the $\textit{Gaia}$ parallax, absolute G magnitude ($M_{G}$) is calculated. Based on this analysis we have estimated the age of 505 stars in our sample. Nearly 106 stars do not have age estimates since majority of them were placed either below the low mass limit of the isochrones or below the main sequence in the CMD \citep{pecaut2013}. Also, it is to be noted that we are not considering the BP/RP excess factor in the sample selection. These factors may account for the non-estimation of ages for the 17$\%$ of our sample.

\begin{figure}[h]
\begin{center}
\includegraphics[width = 1\columnwidth]{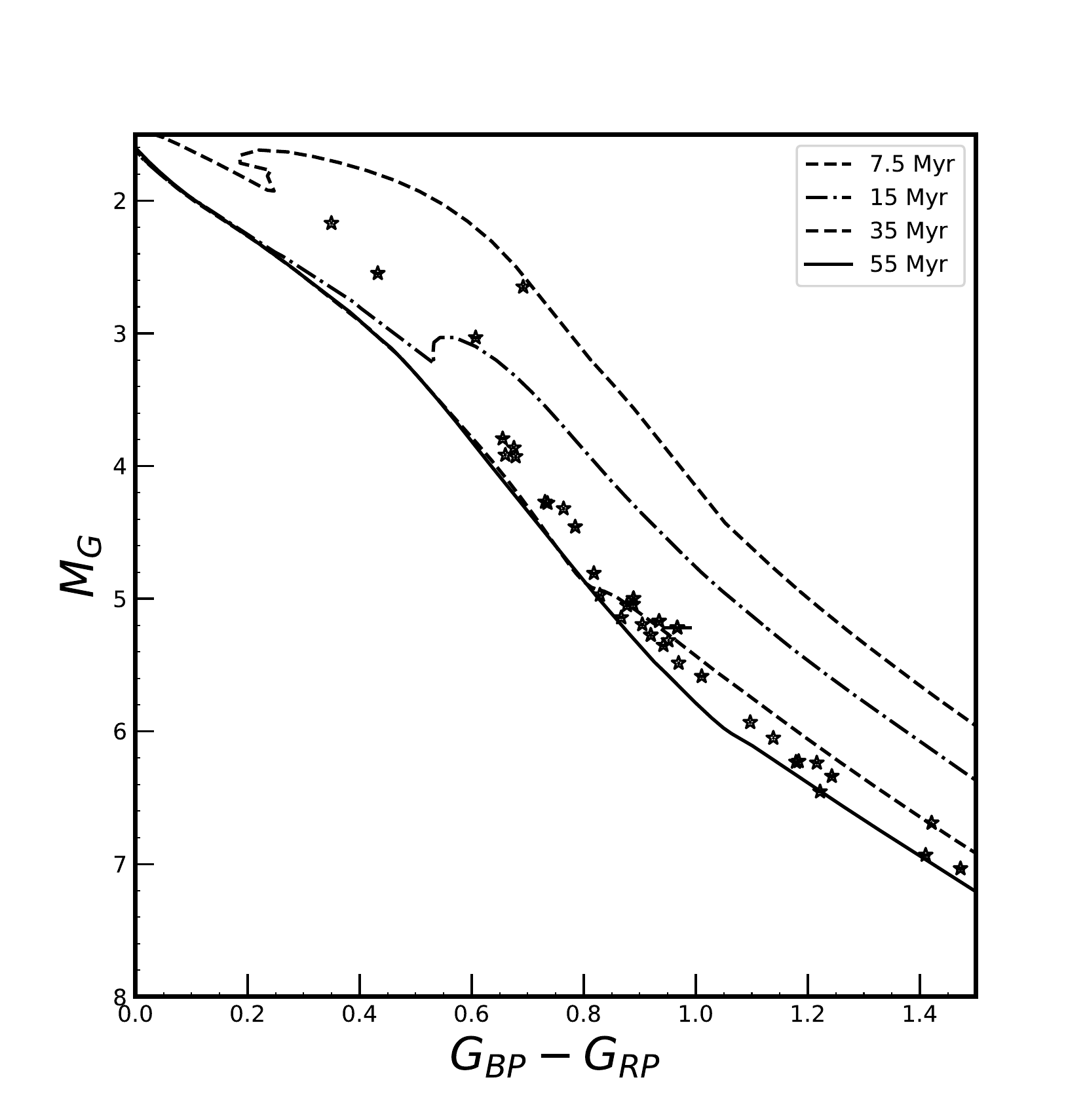}
\caption{Color-magnitude diagram of AB Dor moving group overplotted with MIST isochrones in the age range 7.5 to 55 Myr. From the figure it is evident that the members of a particular moving group shows a range in age.}
\label{fig:CMD}
\end{center}
\end{figure}

The stars belonging to a particular moving group is included in the CMD where isochrones belonging to `Modules for Experiments in Stellar Astrophysics (MESA) Isochrones and Stellar Tracks' (MIST) are plotted. MIST provides stellar evolutionary models for different ages, masses and metallicities \citep{choi2016, Dotter2016}. Isochrones using $\textit{Gaia}$ DR2 data are also included in the latest update of MIST. We have overplotted a grid of MIST isochrones over the CMD and found the closet two isochrones and took the average of the two as the age. The {\it Gaia} CMD ($M_G$ vs $G_{BP}$--$G_{RP}$) of all the members in the AB Dor moving group is shown in Figure \ref{fig:CMD}. Although most of the stars are found to be in the age range 30--50 Myr, a few lower age candidates are also visible. Rather than a single age, a spread in age is visible for all the moving groups. A comparison between the present age estimates and those mentioned in literature is given in Table 3. For most of the moving groups our age estimates agree with those identified from previous studies. However, we found moving groups such as Carina-near to be younger. This may be due to different age estimation methods employed in previous studies. 

\begin{figure*}[t!]
\begin{center}
\includegraphics[width=1\textwidth]{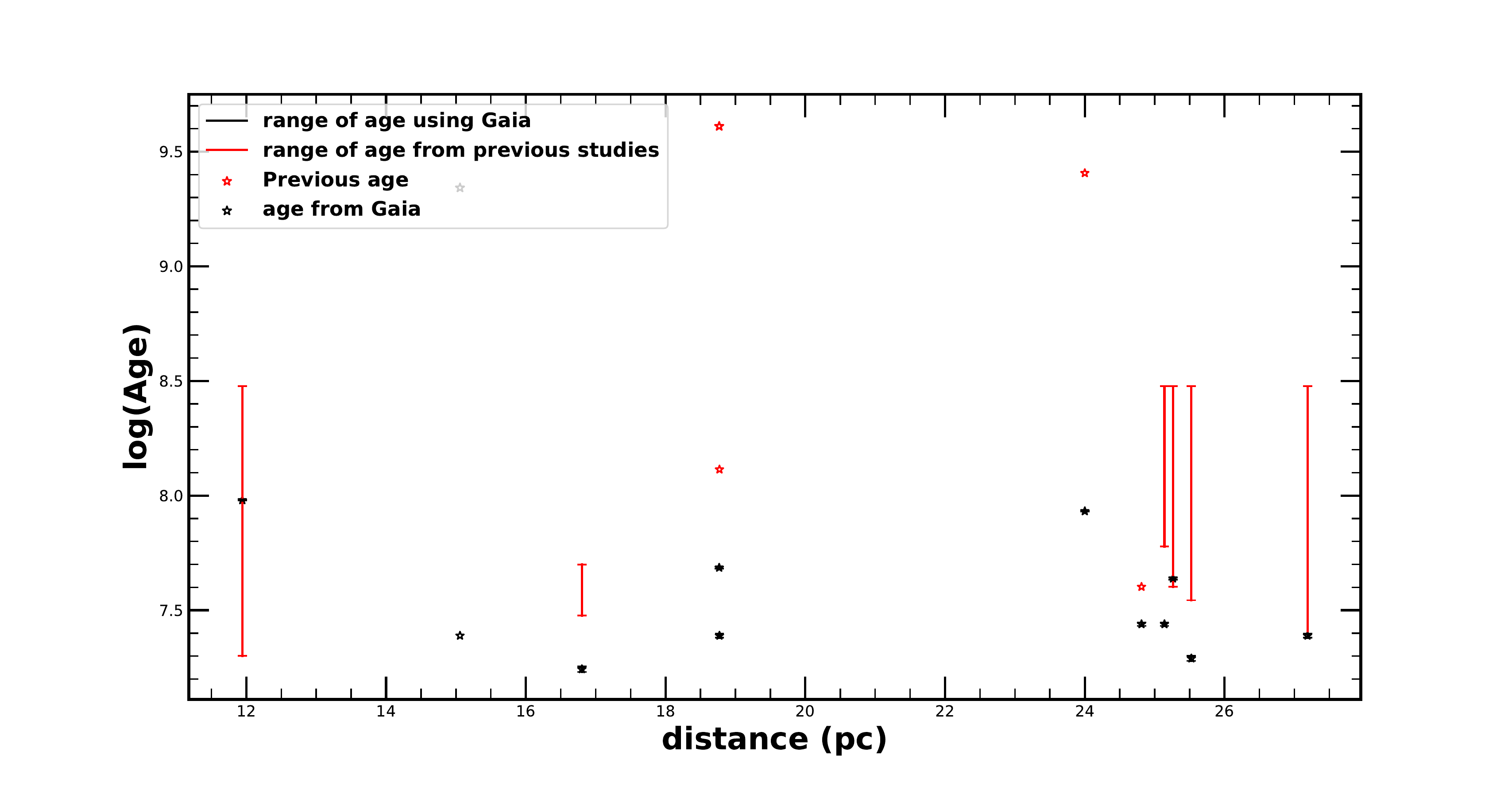}
\caption{Figure represents the comparison between present and previous age estimates of AB Dor moving group members. Red stars and lines represents the age range proposed by previous studies and black stars and lines represent the age range calculated using {\it Gaia} DR2 data. For majority of the stars, our age estimates are in good agreement with previous studies.}
\label{fig:Age}
\end{center}
\end{figure*}

The new age estimates of all the 505 stars, belonging to thirteen moving groups are listed in Table 2. As an example, we discuss the age estimation of 123 members belonging to AB Dor moving group. A comparison of our estimates with those listed in CSNYMGS is shown in Figure \ref{fig:Age}. It is seen that for $\sim$73\% of stars, our new age estimates are considerably near or in the range predicted by previous studies. However, large variations occur for three stars, whose literature values are adopted from \citet{Maldonado2010}. Instead, if we adopt the age of these stars from other works, it matches with our estimates. For example, in the case of G112-035, age mentioned in \citep{Maldonado2010} is 2200 Myr whereas that listed in \cite{Lafreni2007} is 150--300 Myr. This suggests that the estimation of the age of a star is heavily dependent on the techniques and assumptions adopted for estimating the age \citep[e.g,][]{Bell2012,Bell2013}. Since we used {\it Gaia} CMD for the age analysis, the uncertainties in age estimation due to the use of data from diverse resources is mitigated. Hence, the present age estimates are more reliable and can be used for future work. The stellar parameters of moving groups estimated from this work is listed in Table 4.

\section{Conclusions}

We analyzed the membership probability of 1429 moving group candidates listed in `The Catalog of Suspected Nearby Young Moving Group Stars'. {\it Gaia} DR2 parallax and proper motion values are only available for 890 stars. Radial velocity is available for 250 stars, for which {\it U, V, W} space velocity values are estimated. The present work was envisaged to determine the membership probability of stars which were identified as moving group members in CSNYMGS. For this purpose we employed {\it Gaia} DR2 data of moving group stars on a LACEwING framework to produce a new list of moving group members. Based on this analysis we provide a new catalog of moving group members along with the distance, age and space velocity estimates. The main results from this study are summarized below. 

\begin{itemize}
    
    \item We reconfirm that {\it Gaia} DR2 serves as a resource for precise astrometric and photometric measurements of stars. The distance estimated for various moving group candidates were found to be more accurate than previous estimates. Also, the proper motion values from $\textit{Gaia}$ DR2 are found to have minimal errors. This will provide better accuracy in assessing the membership probability of moving group candidates.   
    
    \item The usage of XYZ and UVW ellipses makes LACEwING one of the precise algorithm to determine the moving group membership. The re-analysis of membership probability with the $\textit{Gaia}$ DR2 data using LACEwING algorithm results in a more accurate membership probability for each star. Out of 890 stars listed as moving group stars in CSNYMGS, we found that 279 do not belong to any of the known thirteen moving groups. This needs to be investigated in future studies.    
    
    \item The parallax, proper motion and radial velocity measurements of stars in moving groups are available in $\textit{Gaia}$ DR2. These parameters serve as the input for {\it U, V, W} space velocity estimates. Previous studies had used these input parameters from different sources. This can make the velocity estimates prone to large errors and thereby affecting the moving group membership. With $\textit{Gaia}$ DR2, this issue is solved considerably and hence the space velocity estimates from this work can be used for future studies. 
    
    \item  The young nature of stars are confirmed from the analysis of {\it Gaia} CMD. The age of moving group members ranges from 1.15 Myr to 60.5 Myr. It may be noted that the age estimation using isochrones has systematic uncertainties. However, due to the availability of accurate distance and photometry from {\it Gaia} DR2, we provide the best possible age estimation using isochrone fitting. Also, since we used $G$, $G_{BP}$ and $G_{RP}$ magnitudes in constructing the CMD, large errors in age estimation due to the use of magnitudes from diverse resources is mitigated. 
\end{itemize}

 We believe that studies on moving group can get enriched with the introduction of techniques such as machine learning for estimating the membership probability of young stars. We believe that future data releases of {\it Gaia} can provide precise space velocities of more moving group members and thereby account for incompleteness of data.

\acknowledgments

This work has made use of data from the European Space Agency (ESA) mission Gaia, processed by the Gaia Data Processing and Analysis Consortium (DPAC). Funding for the DPAC has been provided by national institutions, in particular, the institutions participating in the Gaia Multilateral Agreement. Also, this research has made use of the VizieR catalogue access tool, CDS, Strasbourg, France (DOI: 10.26093/cds/vizier). The original description of the VizieR service was published in A$\&$AS 143, 23  
\software{LACEwING \citep{Riedel2016}}

\bibliography{bibtex}

\begin{thebibliography}{}
\expandafter\ifx\csname natexlab\endcsname\relax\def\natexlab#1{#1}\fi
\providecommand{\url}[1]{\href{#1}{#1}}

\bibitem[{{Antoja} {et~al.}(2010){Antoja}, {Figueras}, {Torra}, {Valenzuela},
  \& {Pichardo}}]{Antoja2010}
{Antoja}, T., {Figueras}, F., {Torra}, J., {Valenzuela}, O., \& {Pichardo}, B.
  2010, LNEA, 4, 13

\bibitem[{{Asiain} {et~al.}(1999){Asiain}, {Figueras}, \& {Torra}}]{Asiain1999}
{Asiain}, R., {Figueras}, F., \& {Torra}, J. 1999, \aap, 350, 434

\bibitem[{{Bailey} {et~al.}(2012){Bailey}, {White}, {Blake}, {Charbonneau},
  {Barman}, {Tanner}, \& {Torres}}]{Bailey2012}
{Bailey}, III, J.~I., {White}, R.~J., {Blake}, C.~H., {et~al.} 2012, \apj, 749,
  16

\bibitem[{{Barenfeld} {et~al.}(2013){Barenfeld}, {Bubar}, {Mamajek}, \&
  {Young}}]{Barenfeld2013}
{Barenfeld}, S.~A., {Bubar}, E.~J., {Mamajek}, E.~E., \& {Young}, P.~A. 2013,
  \apj, 766, 6

\bibitem[{{Barrado y Navascu{\'e}s} {et~al.}(2004){Barrado y Navascu{\'e}s},
  {Stauffer}, \& {Jayawardhana}}]{Barrado2004}
{Barrado y Navascu{\'e}s}, D., {Stauffer}, J.~R., \& {Jayawardhana}, R. 2004,
  \apj, 614, 386

\bibitem[{{Barrado y Navascu{\'e}s} {et~al.}(1999){Barrado y Navascu{\'e}s},
  {Stauffer}, {Song}, \& {Caillault}}]{Barrado1999}
{Barrado y Navascu{\'e}s}, D., {Stauffer}, J.~R., {Song}, I., \& {Caillault},
  J.~P. 1999, \apjl, 520, L123

\bibitem[{{Bell} {et~al.}(2015){Bell}, {Mamajek}, \& {Naylor}}]{Bell2015}
{Bell}, C. P.~M., {Mamajek}, E.~E., \& {Naylor}, T. 2015, \mnras, 454, 593

\bibitem[{{Bell} {et~al.}(2012){Bell}, {Naylor}, {Mayne}, {Jeffries}, \&
  {Littlefair}}]{Bell2012}
{Bell}, C.~P.~M., {Naylor}, T., {Mayne}, N.~J., {Jeffries}, R.~D., \&
  {Littlefair}, S.~P. 2012, \mnras, 424, 3178

\bibitem[{{Bell} {et~al.}(2013){Bell}, {Naylor}, {Mayne}, {Jeffries}, \&
  {Littlefair}}]{Bell2013}
---. 2013, \mnras, 434, 806

\bibitem[{{Bell} {et~al.}(2014){Bell}, {Rees}, {Naylor}, {Mayne}, {Jeffries},
  {Mamajek}, \& {Rowe}}]{Bell2014}
{Bell}, C.~P.~M., {Rees}, J.~M., {Naylor}, T., {et~al.} 2014, \mnras, 445, 3496

\bibitem[{{Binks} \& {Jeffries}(2017)}]{Binks2017}
{Binks}, A.~S., \& {Jeffries}, R.~D. 2017, \mnras, 469, 579

\bibitem[{{Bowler} {et~al.}(2013){Bowler}, {Liu}, {Shkolnik}, \&
  {Dupuy}}]{Bowler2013}
{Bowler}, B.~P., {Liu}, M.~C., {Shkolnik}, E.~L., \& {Dupuy}, T.~J. 2013, \apj,
  774, 55

\bibitem[{{Bowler} {et~al.}(2017){Bowler}, {Liu}, {Mawet}, {Ngo}, {Malo},
  {Mace}, {McLane}, {Lu}, {Tristan}, {Hinkley}, {Hillenbrand}, {Shkolnik},
  {Benneke}, \& {Best}}]{Bowler2017}
{Bowler}, B.~P., {Liu}, M.~C., {Mawet}, D., {et~al.} 2017, \aj, 153, 18

\bibitem[{{Chauvin} {et~al.}(2004){Chauvin}, {Lagrange}, {Dumas}, {Zuckerman},
  {Mouillet}, {Song}, {Beuzit}, \& {Lowrance}}]{Chauvin2004}
{Chauvin}, G., {Lagrange}, A.~M., {Dumas}, C., {et~al.} 2004, \aap, 425, L29

\bibitem[{{Chauvin} {et~al.}(2005){Chauvin}, {Lagrange}, {Dumas}, {Zuckerman},
  {Mouillet}, {Song}, {Beuzit}, \& {Lowrance}}]{Chauvin2005}
---. 2005, \aap, 438, L25

\bibitem[{{Choi} {et~al.}(2016){Choi}, {Dotter}, {Conroy}, {Cantiello},
  {Paxton}, \& {Johnson}}]{choi2016}
{Choi}, J., {Dotter}, A., {Conroy}, C., {et~al.} 2016, \apj, 823, 102

\bibitem[{{Cropper} {et~al.}(2018){Cropper}, {Katz}, {Sartoretti}, {Prusti},
  {de Bruijne}, {Chassat}, {Charvet}, {Boyadjian}, {Perryman}, \&
  {Sarri}}]{Cropper2018}
{Cropper}, M., {Katz}, D., {Sartoretti}, P., {et~al.} 2018, \aap, 616, A5

\bibitem[{{Crundall} {et~al.}(2019){Crundall}, {Ireland}, {Krumholz},
  {Federrath}, {{\v{Z}}erjal}, \& {Hansen}}]{Chronostar2019}
{Crundall}, T.~D., {Ireland}, M.~J., {Krumholz}, M.~R., {et~al.} 2019, \mnras,
  489, 3625

\bibitem[{{da Silva} {et~al.}(2009){da Silva}, {Torres}, {de La Reza}, {Quast},
  {Melo}, \& {Sterzik}}]{dasilva2009}
{da Silva}, L., {Torres}, C.~A.~O., {de La Reza}, R., {et~al.} 2009, \aap, 508,
  833

\bibitem[{{de la Reza} {et~al.}(1989){de la Reza}, {Torres}, {Quast},
  {Castilho}, \& {Vieira}}]{delaReza1989}
{de la Reza}, R., {Torres}, C. A.~O., {Quast}, G., {Castilho}, B.~V., \&
  {Vieira}, G.~L. 1989, \apjl, 343, L61

\bibitem[{{De Silva} {et~al.}(2013){De Silva}, {D'Orazi}, {Melo}, {Torres},
  {Gieles}, {Quast}, \& {Sterzik}}]{desilva2013}
{De Silva}, G.~M., {D'Orazi}, V., {Melo}, C., {et~al.} 2013, MNRAS, 431, 1005

\bibitem[{{Dias} {et~al.}(2002){Dias}, {Alessi}, {Moitinho}, \&
  {L{\'e}pine}}]{Dias2002}
{Dias}, W.~S., {Alessi}, B.~S., {Moitinho}, A., \& {L{\'e}pine}, J.~R.~D. 2002,
  \aap, 389, 871

\bibitem[{{Dotter}(2016)}]{Dotter2016}
{Dotter}, A. 2016, \apjs, 222, 8

\bibitem[{{Dupuy} {et~al.}(2018){Dupuy}, {Liu}, {Allers}, {Biller}, {Kratter},
  {Mann}, {Shkolnik}, {Kraus}, \& {Best}}]{Dupuy2018}
{Dupuy}, T.~J., {Liu}, M.~C., {Allers}, K.~N., {et~al.} 2018, \aj, 156, 57

\bibitem[{{Eggen}(1996)}]{Eggen1996}
{Eggen}, O.~J. 1996, \aj, 112, 1595

\bibitem[{{Elliott} {et~al.}(2014){Elliott}, {Bayo}, {Melo}, {Torres},
  {Sterzik}, \& {Quast}}]{Elliot2014}
{Elliott}, P., {Bayo}, A., {Melo}, C.~H.~F., {et~al.} 2014, \aap, 568, A26

\bibitem[{{Elliott} {et~al.}(2016){Elliott}, {Bayo}, {Melo}, {Torres},
  {Sterzik}, {Quast}, {Montes}, \& {Brahm}}]{Elliot2016}
---. 2016, \aap, 590, A13

\bibitem[{{Elliott} {et~al.}(2015){Elliott}, {Hu{\'e}lamo}, {Bouy}, {Bayo},
  {Melo}, {Torres}, {Sterzik}, {Quast}, {Chauvin}, \& {Barrado}}]{Elliot2015}
{Elliott}, P., {Hu{\'e}lamo}, N., {Bouy}, H., {et~al.} 2015, \aap, 580, A88

\bibitem[{{Evans} {et~al.}(2018){Evans}, {Riello}, {De Angeli}, {Carrasco},
  {Montegriffo}, {Fabricius}, {Jordi}, {Palaversa}, {Diener}, {Busso},
  {Cacciari}, {van Leeuwen}, {Burgess}, {Davidson}, {Harrison}, {Hodgkin},
  {Pancino}, {Richards}, {Altavilla}, {Balaguer-N{\'u}{\~n}ez}, {Barstow},
  {Bellazzini}, {Brown}, {Castellani}, {Cocozza}, {De Luise}, {Delgado},
  {Ducourant}, {Galleti}, {Gilmore}, {Giuffrida}, {Holl}, {Kewley}, {Koposov},
  {Marinoni}, {Marrese}, {Osborne}, {Piersimoni}, {Portell}, {Pulone},
  {Ragaini}, {Sanna}, {Terrett}, {Walton}, {Wevers}, \&
  {Wyrzykowski}}]{Evans2018}
{Evans}, D.~W., {Riello}, M., {De Angeli}, F., {et~al.} 2018, \aap, 616, A4

\bibitem[{{Faherty} {et~al.}(2018){Faherty}, {Bochanski}, {Gagn{\'e}},
  {Nelson}, {Coker}, {Smithka}, {Desir}, \& {Vasquez}}]{Faherty2018}
{Faherty}, J.~K., {Bochanski}, J.~J., {Gagn{\'e}}, J., {et~al.} 2018, \apj,
  863, 91

\bibitem[{{Famaey} {et~al.}(2008){Famaey}, {Siebert}, \&
  {Jorissen}}]{Famaey2008}
{Famaey}, B., {Siebert}, A., \& {Jorissen}, A. 2008, \aap, 483, 453

\bibitem[{{Fuhrmann}(2004)}]{Fuhrmann2004}
{Fuhrmann}, K. 2004, Astronomische Nachrichten, 325, 3

\bibitem[{{Gagn{\'e}} \& {Faherty}(2018)}]{GagneDR22018ApJ...862..138G}
{Gagn{\'e}}, J., \& {Faherty}, J.~K. 2018, \apj, 862, 138

\bibitem[{{Gagn{\'e}} {et~al.}(2014){Gagn{\'e}}, {Lafreni{\`e}re}, {Doyon},
  {Malo}, \& {Artigau}}]{Gagne2014}
{Gagn{\'e}}, J., {Lafreni{\`e}re}, D., {Doyon}, R., {Malo}, L., \& {Artigau},
  {\'E}. 2014, \apj, 783, 121

\bibitem[{{Gagn{\'e}} {et~al.}(2018{\natexlab{a}}){Gagn{\'e}}, {Roy-Loubier},
  {Faherty}, {Doyon}, \& {Malo}}]{Gagne2018}
{Gagn{\'e}}, J., {Roy-Loubier}, O., {Faherty}, J.~K., {Doyon}, R., \& {Malo},
  L. 2018{\natexlab{a}}, ApJ, 860, 43

\bibitem[{{Gagn{\'e}} {et~al.}(2018{\natexlab{b}}){Gagn{\'e}}, {Mamajek},
  {Malo}, {Riedel}, {Rodriguez}, {Lafreni{\`e}re}, {Faherty}, {Roy-Loubier},
  {Pueyo}, {Robin}, \& {Doyon}}]{BANSIG2018}
{Gagn{\'e}}, J., {Mamajek}, E.~E., {Malo}, L., {et~al.} 2018{\natexlab{b}},
  \apj, 856, 23

\bibitem[{{Gaia Collaboration} {et~al.}(2016){Gaia Collaboration}, {Brown},
  {Vallenari}, {Prusti}, {de Bruijne}, {Mignard}, {Drimmel}, {Babusiaux},
  {Bailer-Jones}, {Bastian}, \& et~al.}]{Gaia2016}
{Gaia Collaboration}, {Brown}, A.~G.~A., {Vallenari}, A., {et~al.} 2016, \aap,
  595, A2

\bibitem[{{Gaia Collaboration} {et~al.}(2018{\natexlab{a}}){Gaia
  Collaboration}, {Brown}, {Vallenari}, {Prusti}, {de Bruijne}, {Babusiaux},
  {Bailer-Jones}, {Biermann}, {Evans}, {Eyer}, \& et~al.}]{dr2018}
---. 2018{\natexlab{a}}, \aap, 616, A1

\bibitem[{{Gaia Collaboration} {et~al.}(2018{\natexlab{b}}){Gaia
  Collaboration}, {Babusiaux}, {van Leeuwen}, {Barstow}, {Jordi}, {Vallenari},
  {Bossini}, {Bressan}, {Cantat-Gaudin}, {van Leeuwen}, \& et~al.}]{GaiaHR2018}
{Gaia Collaboration}, {Babusiaux}, C., {van Leeuwen}, F., {et~al.}
  2018{\natexlab{b}}, \aap, 616, A10

\bibitem[{{Gaidos}(1998)}]{Gaidos1998}
{Gaidos}, E.~J. 1998, \pasp, 110, 1259

\bibitem[{{Henden} {et~al.}(2016){Henden}, {Templeton}, {Terrell}, {Smith},
  {Levine}, \& {Welch}}]{Henden2016}
{Henden}, A.~A., {Templeton}, M., {Terrell}, D., {et~al.} 2016, VizieR Online
  Data Catalog, 2336

\bibitem[{{Hoogerwerf} \& {Aguilar}(1999)}]{Hoogerwerf1999}
{Hoogerwerf}, R., \& {Aguilar}, L.~A. 1999, \mnras, 306, 394

\bibitem[{{Jeffries}(1995)}]{Jeffries1995}
{Jeffries}, R.~D. 1995, \mnras, 273, 559

\bibitem[{{Johnson} \& {Soderblom}(1987)}]{Johnson1987}
{Johnson}, D.~R.~H., \& {Soderblom}, D.~R. 1987, AJ, 93, 864

\bibitem[{{Jones} {et~al.}(2015){Jones}, {White}, {Boyajian}, {Schaefer},
  {Baines}, {Ireland}, {Patience}, {ten Brummelaar}, {McAlister}, {Ridgway},
  {Sturmann}, {Sturmann}, {Turner}, {Farrington}, \& {Goldfinger}}]{Jones2015}
{Jones}, J., {White}, R.~J., {Boyajian}, T., {et~al.} 2015, \apj, 813, 58

\bibitem[{{King} {et~al.}(2003){King}, {Villarreal}, {Soderblom}, {Gulliver},
  \& {Adelman}}]{King2003}
{King}, J.~R., {Villarreal}, A.~R., {Soderblom}, D.~R., {Gulliver}, A.~F., \&
  {Adelman}, S.~J. 2003, \aj, 125, 1980

\bibitem[{{Koen} {et~al.}(2010){Koen}, {Kilkenny}, {van Wyk}, \&
  {Marang}}]{Koen2010}
{Koen}, C., {Kilkenny}, D., {van Wyk}, F., \& {Marang}, F. 2010, \mnras, 403,
  1949

\bibitem[{{Kraus} {et~al.}(2014){Kraus}, {Shkolnik}, {Allers}, \&
  {Liu}}]{Kraus2014}
{Kraus}, A.~L., {Shkolnik}, E.~L., {Allers}, K.~N., \& {Liu}, M.~C. 2014, \aj,
  147, 146

\bibitem[{{Lafreni{\`e}re} {et~al.}(2007){Lafreni{\`e}re}, {Doyon}, {Marois},
  {Nadeau}, {Oppenheimer}, {Roche}, {Rigaut}, {Graham}, {Jayawardhana},
  {Johnstone}, {Kalas}, {Macintosh}, \& {Racine}}]{Lafreni2007}
{Lafreni{\`e}re}, D., {Doyon}, R., {Marois}, C., {et~al.} 2007, \apj, 670, 1367

\bibitem[{{Lagrange} {et~al.}(2010){Lagrange}, {Bonnefoy}, {Chauvin}, {Apai},
  {Ehrenreich}, {Boccaletti}, {Gratadour}, {Rouan}, {Mouillet}, \&
  {Lacour}}]{Lagrange2010}
{Lagrange}, A.~M., {Bonnefoy}, M., {Chauvin}, G., {et~al.} 2010, Science, 329,
  57

\bibitem[{{Lindegren}(2018)}]{Lindegren2018}
{Lindegren}, L. 2018, Gaia Technical Note: GAIA-C3-TN-LU-LL-124-01

\bibitem[{{Makarov}(2007)}]{Makarov2007}
{Makarov}, V.~V. 2007, \apjs, 169, 105

\bibitem[{{Maldonado} {et~al.}(2010){Maldonado}, {Mart{\'{\i}}nez-Arn{\'a}iz},
  {Eiroa}, {Montes}, \& {Montesinos}}]{Maldonado2010}
{Maldonado}, J., {Mart{\'{\i}}nez-Arn{\'a}iz}, R.~M., {Eiroa}, C., {Montes},
  D., \& {Montesinos}, B. 2010, \aap, 521, A12

\bibitem[{{Malo} {et~al.}(2014){Malo}, {Doyon}, {Feiden}, {Albert},
  {Lafreni{\`e}re}, {Artigau}, {Gagn{\'e}}, \& {Riedel}}]{Malo2014}
{Malo}, L., {Doyon}, R., {Feiden}, G.~A., {et~al.} 2014, \apj, 792, 37

\bibitem[{{Malo} {et~al.}(2013){Malo}, {Doyon}, {Lafreni{\`e}re}, {Artigau},
  {Gagn{\'e}}, {Baron}, \& {Riedel}}]{Malo2013}
{Malo}, L., {Doyon}, R., {Lafreni{\`e}re}, D., {et~al.} 2013, \apj, 762, 88

\bibitem[{{Mamajek}(2005)}]{Mamajek2005}
{Mamajek}, E.~E. 2005, \apj, 634, 1385

\bibitem[{{Mamajek} {et~al.}(2000){Mamajek}, {Lawson}, \&
  {Feigelson}}]{Mamajek2000}
{Mamajek}, E.~E., {Lawson}, W.~A., \& {Feigelson}, E.~D. 2000, \apj, 544, 356

\bibitem[{{Marrese} {et~al.}(2019){Marrese}, {Marinoni}, {Fabrizio}, \&
  {Altavilla}}]{Marrese2019}
{Marrese}, P.~M., {Marinoni}, S., {Fabrizio}, M., \& {Altavilla}, G. 2019,
  \aap, 621, A144

\bibitem[{{Mathew} {et~al.}(2017){Mathew}, {Manoj}, {Bhatt}, {Sahu},
  {Maheswar}, \& {Muneer}}]{Mathew2017}
{Mathew}, B., {Manoj}, P., {Bhatt}, B.~C., {et~al.} 2017, \aj, 153, 225

\bibitem[{{Meingast} \& {Alves}(2019)}]{Meingast2019_1}
{Meingast}, S., \& {Alves}, J. 2019, \aap, 621, L3

\bibitem[{{Meingast} {et~al.}(2019){Meingast}, {Alves}, \&
  {F{\"u}rnkranz}}]{Meingast2019}
{Meingast}, S., {Alves}, J., \& {F{\"u}rnkranz}, V. 2019, \aap, 622, L13

\bibitem[{{Messina} {et~al.}(2010){Messina}, {Desidera}, {Turatto},
  {Lanzafame}, \& {Guinan}}]{Messina}
{Messina}, S., {Desidera}, S., {Turatto}, M., {Lanzafame}, A.~C., \& {Guinan},
  E.~F. 2010, \aap, 520, A15

\bibitem[{{Montes} {et~al.}(2001){Montes}, {L{\'o}pez-Santiago}, {G{\'a}lvez},
  {Fern{\'a}ndez-Figueroa}, {De Castro}, \& {Cornide}}]{montes2001}
{Montes}, D., {L{\'o}pez-Santiago}, J., {G{\'a}lvez}, M.~C., {et~al.} 2001,
  \mnras, 328, 45

\bibitem[{{Murphy} \& {Lawson}(2015)}]{Murphy2015}
{Murphy}, S.~J., \& {Lawson}, W.~A. 2015, MNRAS, 447, 1267

\bibitem[{{Murphy} {et~al.}(2013){Murphy}, {Lawson}, \& {Bessell}}]{Murphy2013}
{Murphy}, S.~J., {Lawson}, W.~A., \& {Bessell}, M.~S. 2013, \mnras, 435, 1325

\bibitem[{{Ortega} {et~al.}(2002){Ortega}, {de la Reza}, {Jilinski}, \&
  {Bazzanella}}]{ortega2002}
{Ortega}, V.~G., {de la Reza}, R., {Jilinski}, E., \& {Bazzanella}, B. 2002,
  \apjl, 575, L75

\bibitem[{{Ortega} {et~al.}(2004){Ortega}, {de la Reza}, {Jilinski}, \&
  {Bazzanella}}]{ortega2004}
---. 2004, \apj, 609, 243

\bibitem[{{Pecaut} \& {Mamajek}(2013)}]{pecaut2013}
{Pecaut}, M.~J., \& {Mamajek}, E.~E. 2013, \apjs, 208, 9

\bibitem[{{P{\"o}hnl} \& {Paunzen}(2010)}]{Pohnl2010}
{P{\"o}hnl}, H., \& {Paunzen}, E. 2010, \aap, 514, A81

\bibitem[{{Ramya} {et~al.}(2012){Ramya}, {Reddy}, \& {Lambert}}]{Ramya2012}
{Ramya}, P., {Reddy}, B.~E., \& {Lambert}, D.~L. 2012, \mnras, 425, 3188

\bibitem[{{Rebull} {et~al.}(2008){Rebull}, {Stapelfeldt}, {Werner}, {Mannings},
  {Chen}, {Stauffer}, {Smith}, {Song}, {Hines}, \& {Low}}]{Rebull2008}
{Rebull}, L.~M., {Stapelfeldt}, K.~R., {Werner}, M.~W., {et~al.} 2008, \apj,
  681, 1484

\bibitem[{{Riedel}(2016)}]{Riedel2016}
{Riedel}, A.~R. 2016, in IAU Symposium, Vol. 314, Young Stars \&amp; Planets
  Near the Sun, ed. J.~H. {Kastner}, B.~{Stelzer}, \& S.~A. {Metchev}, 33--36

\bibitem[{{Riedel} {et~al.}(2017){Riedel}, {Blunt}, {Lambrides}, {Rice},
  {Cruz}, \& {Faherty}}]{Riedel2017}
{Riedel}, A.~R., {Blunt}, S.~C., {Lambrides}, E.~L., {et~al.} 2017, AJ, 153, 95

\bibitem[{{Rodriguez} {et~al.}(2011){Rodriguez}, {Bessell}, {Zuckerman}, \&
  {Kastner}}]{Rodriguez2011}
{Rodriguez}, D.~R., {Bessell}, M.~S., {Zuckerman}, B., \& {Kastner}, J.~H.
  2011, \apj, 727, 62

\bibitem[{{Rodriguez} {et~al.}(2015){Rodriguez}, {van der Plas}, {Kastner},
  {Schneider}, {Faherty}, {Mardones}, {Mohanty}, \& {Principe}}]{Rodriguez2015}
{Rodriguez}, D.~R., {van der Plas}, G., {Kastner}, J.~H., {et~al.} 2015, \aap,
  582, L5

\bibitem[{{R{\"o}ser} {et~al.}(2019){R{\"o}ser}, {Schilbach}, \&
  {Goldman}}]{Roser2019}
{R{\"o}ser}, S., {Schilbach}, E., \& {Goldman}, B. 2019, \aap, 621, L2

\bibitem[{{Schlieder} {et~al.}(2012){Schlieder}, {L{\'e}pine}, \&
  {Simon}}]{Schlieder2012}
{Schlieder}, J.~E., {L{\'e}pine}, S., \& {Simon}, M. 2012, \aj, 144, 109

\bibitem[{{Schneider} {et~al.}(2019){Schneider}, {Shkolnik}, {Allers}, {Kraus},
  {Liu}, {Weinberger}, \& {Flagg}}]{Schneider2019A}
{Schneider}, A.~C., {Shkolnik}, E.~L., {Allers}, K.~N., {et~al.} 2019, \aj,
  157, 234

\bibitem[{{Shkolnik} {et~al.}(2017){Shkolnik}, {Allers}, {Kraus}, {Liu}, \&
  {Flagg}}]{Shkolnik2017}
{Shkolnik}, E.~L., {Allers}, K.~N., {Kraus}, A.~L., {Liu}, M.~C., \& {Flagg},
  L. 2017, \aj, 154, 69

\bibitem[{{Shkolnik} {et~al.}(2012){Shkolnik}, {Anglada-Escud{\'e}}, {Liu},
  {Bowler}, {Weinberger}, {Boss}, {Reid}, \& {Tamura}}]{Shkolnik2012}
{Shkolnik}, E.~L., {Anglada-Escud{\'e}}, G., {Liu}, M.~C., {et~al.} 2012, \apj,
  758, 56

\bibitem[{{Soderblom} {et~al.}(2014){Soderblom}, {Hillenbrand}, {Jeffries},
  {Mamajek}, \& {Naylor}}]{Soderblom2014}
{Soderblom}, D.~R., {Hillenbrand}, L.~A., {Jeffries}, R.~D., {Mamajek}, E.~E.,
  \& {Naylor}, T. 2014, Protostars and Planets VI, 219

\bibitem[{{Song} {et~al.}(2003){Song}, {Zuckerman}, \& {Bessell}}]{Song2003}
{Song}, I., {Zuckerman}, B., \& {Bessell}, M.~S. 2003, \apj, 599, 342

\bibitem[{{Torres} {et~al.}(2000){Torres}, {da Silva}, {Quast}, {de la Reza},
  \& {Jilinski}}]{Torres2000}
{Torres}, C.~A.~O., {da Silva}, L., {Quast}, G.~R., {de la Reza}, R., \&
  {Jilinski}, E. 2000, \aj, 120, 1410

\bibitem[{{Torres} {et~al.}(2006){Torres}, {Quast}, {da Silva}, {de La Reza},
  {Melo}, \& {Sterzik}}]{Torres2006}
{Torres}, C.~A.~O., {Quast}, G.~R., {da Silva}, L., {et~al.} 2006, \aap, 460,
  695

\bibitem[{{Torres} {et~al.}(2003){Torres}, {Quast}, {de La Reza}, {da Silva},
  {Melo}, \& {Sterzik}}]{Torres2003}
{Torres}, C. A.~O., {Quast}, G.~R., {de La Reza}, R., {et~al.} 2003, in
  Astrophysics and Space Science Library, Vol. 299, Astrophysics and Space
  Science Library, ed. J.~{L{\'e}pine} \& J.~{Gregorio-Hetem}, 83

\bibitem[{{Torres} {et~al.}(2008){Torres}, {Quast}, {Melo}, \&
  {Sterzik}}]{Torres2008}
{Torres}, C.~A.~O., {Quast}, G.~R., {Melo}, C.~H.~F., \& {Sterzik}, M.~F. 2008,
  {Young Nearby Loose Associations}, ed. B.~{Reipurth}, 757

\bibitem[{{van Leeuwen}(2007)}]{hipp2007}
{van Leeuwen}, F. 2007, \aap, 474, 653

\bibitem[{{Viana Almeida} {et~al.}(2009){Viana Almeida}, {Santos}, {Melo},
  {Ammler-von Eiff}, {Torres}, {Quast}, {Gameiro}, \& {Sterzik}}]{Viana2009}
{Viana Almeida}, P., {Santos}, N.~C., {Melo}, C., {et~al.} 2009, \aap, 501, 965

\bibitem[{{Weinberger} {et~al.}(2013){Weinberger}, {Anglada-Escud{\'e}}, \&
  {Boss}}]{Weinberger2013}
{Weinberger}, A.~J., {Anglada-Escud{\'e}}, G., \& {Boss}, A.~P. 2013, \apj,
  762, 118

\bibitem[{{Zuckerman}(2001)}]{zuckerman2001}
{Zuckerman}, B. 2001, in Astronomical Society of the Pacific Conference Series,
  Vol. 244, Young Stars Near Earth: Progress and Prospects, ed.
  R.~{Jayawardhana} \& T.~{Greene}, 122

\bibitem[{{Zuckerman} {et~al.}(2006){Zuckerman}, {Bessell}, {Song}, \&
  {Kim}}]{Zuckerman2006}
{Zuckerman}, B., {Bessell}, M.~S., {Song}, I., \& {Kim}, S. 2006, \apjl, 649,
  L115

\bibitem[{{Zuckerman} {et~al.}(2011){Zuckerman}, {Rhee}, {Song}, \&
  {Bessell}}]{Zuckerman2011}
{Zuckerman}, B., {Rhee}, J.~H., {Song}, I., \& {Bessell}, M.~S. 2011, \apj,
  732, 61

\bibitem[{{Zuckerman} \& {Song}(2004)}]{Zuckerman2004}
{Zuckerman}, B., \& {Song}, I. 2004, \araa, 42, 685

\bibitem[{{Zuckerman} {et~al.}(2001){Zuckerman}, {Song}, \&
  {Webb}}]{zuckermanetal.2001}
{Zuckerman}, B., {Song}, I., \& {Webb}, R.~A. 2001, \apj, 559, 388

\end{thebibliography}

\newpage 
\begin{center}
\begin{tiny}
\begin{longtable}{llllclll}
\caption{ The {\it U, V, W} values estimated from $\textit{Gaia}$ DR2 for ABDor moving group members is shown along with those listed in CSNYMGS compiled from previous studies.}\\
\hline
               & \multicolumn{4}{c}{Previous studies (km $s^{-1}$)}                                       & \multicolumn{3}{c}{\textit{Gaia} estimates (km $s^{-1}$)}                             \\
\endfirsthead
\endhead
Name           & \multicolumn{1}{c}{U} & \multicolumn{1}{c}{V} & \multicolumn{1}{c}{W} & Reference & \multicolumn{1}{c}{U} & \multicolumn{1}{c}{V} & \multicolumn{1}{c}{W} \\ \hline
BD-00 632      & -7.9$\pm$0.2          & -28.1$\pm$0.2         & -12$\pm$0.1           & 2         & -7.85$\pm$0.21        & -28$\pm$0.07          & -11.79$\pm$0.15       \\
BD-12 243      & -4.2$\pm$0.2          & -27.2$\pm$1           & -12$\pm$0.4           & 2         & -5.04$\pm$0.04        & -27.6$\pm$0.06        & -13.95$\pm$0.15       \\
BD-15 200      & -7.8$\pm$0.5          & -29$\pm$1.8           & -9.3$\pm$1.2          & 2         & -7.94$\pm$0.06        & -27.56$\pm$0.07       & -10.99$\pm$0.29       \\
BD-19 6489     & -28.72$\pm$0.84       & -17.19$\pm$0.57       & -21.13$\pm$0.48       & 1         & -29.21$\pm$0.07       & -17.36$\pm$0.06       & -20.61$\pm$0.16       \\
BD+00 3243     & -9.76$\pm$0.18        & -26.29$\pm$1.31       & -10.75$\pm$0.13       & 1         & -9.66$\pm$0.13        & -28.07$\pm$0.07       & -10.73$\pm$0.16       \\
BD+01 699      & -5$\pm$1.1            & -27$\pm$1.3           & -14.3$\pm$0.8         & 2         & -5.38$\pm$0.36        & -27.88$\pm$0.09       & -15$\pm$0.25          \\
BD+07 1919A    & -5.53$\pm$4.19        & -32.52$\pm$4.38       & -4.13$\pm$2.55        & 1         & -3.57$\pm$0.31        & -31.53$\pm$0.22       & -5.51$\pm$0.15        \\
BD+07 1919B    & -9.01$\pm$4.02        & -25.29$\pm$3.64       & -0.63$\pm$2.19        & 1         & -3.57$\pm$0.31        & -31.53$\pm$0.22       & -5.51$\pm$0.15        \\
BD+21 418 A    & -5.8$\pm$0.3          & -29.8$\pm$1.9         & -18$\pm$1             & 2         & -5.32$\pm$0.33        & -28.41$\pm$0.13       & -16.36$\pm$0.2        \\
1E 0318.5-19.4 & -12.69$\pm$0.74       & -17.27$\pm$1.66       & -11.85$\pm$1.09       & 1         & -12.95$\pm$0.38       & -22.82$\pm$0.21       & -8.15$\pm$0.6         \\
BD+23 296 A    & -8.6$\pm$0.4          & -31.3$\pm$1.4         & -14.3$\pm$0.7         & 2         & -8.22$\pm$0.16        & -27.41$\pm$0.12       & -12.52$\pm$0.13       \\
BD+23 296 C    & -8.6$\pm$0.4          & -31.3$\pm$1.4         & -14.3$\pm$0.7         & 2         & -8.22$\pm$0.16        & -27.41$\pm$0.12       & -12.52$\pm$0.13       \\
BD+37 604 B    & -8.2$\pm$1.6          & -28.2$\pm$3.5         & -13.7$\pm$2.1         & 2         & -9.53$\pm$1.11        & -27.85$\pm$0.75       & -12.02$\pm$0.46       \\
BD+41 2076     & -10.41$\pm$0.84       & -29.62$\pm$0.85       & -14.16$\pm$1.12       & 1         & -11.09$\pm$0.11       & -29.41$\pm$0.08       & -13.06$\pm$0.15       \\
BD+41 4749     & -4.3$\pm$0.6          & -27.4$\pm$0.6         & -14.6$\pm$1.2         & 2         & -4.65$\pm$0.07        & -27.68$\pm$0.25       & -14.79$\pm$0.09       \\
CD-26 16420 B  & -2.6$\pm$0.7          & -25.2$\pm$1.7         & -15.8$\pm$1.5         & 2         & -1.54$\pm$0.14        & -26.35$\pm$0.09       & -13.65$\pm$0.35       \\
CD-35 2722 A   & -7.5$\pm$5            & -27.5$\pm$0.8         & -14.6$\pm$0.4         & 4         & -7.11$\pm$0.82        & -25.89$\pm$1.59       & -13.65$\pm$0.77       \\
CD-35 2722 B   & -7.5$\pm$5            & -27.5$\pm$0.8         & -14.6$\pm$0.4         & 4         & -7.11$\pm$0.82        & -25.89$\pm$1.59       & -13.65$\pm$0.77       \\
CD-36 15990    & -7.3$\pm$1            & -27.3$\pm$2           & -13.5$\pm$1.6         & 2         & -8.83$\pm$1.53        & -27.53$\pm$0.11       & -11.12$\pm$5.12       \\
CD-39 1951 A   & -7.4$\pm$0.1          & -27.7$\pm$0.2         & -14.8$\pm$0.1         & 2         & -7.86$\pm$0.09        & -27.86$\pm$0.18       & -15.06$\pm$0.13       \\
CD-39 1951 B   & -7.4$\pm$0.1          & -27.7$\pm$0.2         & -14.8$\pm$0.1         & 2         & -7.86$\pm$0.09        & -27.86$\pm$0.18       & -15.06$\pm$0.13       \\
CD-60 1425     & -7.6$\pm$0.1          & -27.7$\pm$0.2         & -14.3$\pm$0.1         & 2         & -7.68$\pm$0.01        & -27.92$\pm$0.19       & -14.29$\pm$0.09       \\
CD-60 7126     & -7.6$\pm$0.8          & -27.3$\pm$1.2         & -11.7$\pm$0.4         & 2         & -5.62$\pm$0.37        & -27.04$\pm$0.17       & -11.7$\pm$0.2         \\
CD-61 1439     & -7.8$\pm$0.2          & -27.3$\pm$0.6         & -14.1$\pm$0.3         & 2         & -7.71$\pm$0.02        & -28.32$\pm$0.67       & -14.37$\pm$0.31       \\
G 112-035      & -12.25$\pm$0.17       & -22.15$\pm$0.27       & -12.5$\pm$0.39        & 3         & -11.41$\pm$0.43       & -21.95$\pm$0.32       & -13.02$\pm$0.12       \\
G 247-013      & -7.92$\pm$0.26        & -23.86$\pm$0.23       & -17.19$\pm$0.22       & 3         & -7.8$\pm$0.1          & -24.02$\pm$0.09       & -17.15$\pm$0.03       \\
G 267-001      & -7.8$\pm$0.3          & -27.6$\pm$0.5         & -12.5$\pm$0.5         & 2         & -7.49$\pm$0.06        & -27.75$\pm$0.06       & -12.91$\pm$0.21       \\
HIP 114066     & -6.1$\pm$0.7          & -27$\pm$0.8           & -15.5$\pm$0.6         & 2         & -5.73$\pm$0.84        & -27.71$\pm$2.07       & -15.33$\pm$0.14       \\
HIP 15442      & -25.57$\pm$0.22       & -30.64$\pm$0.57       & -6.09$\pm$0.24        & 3         & -24.4$\pm$0.14        & -29.18$\pm$0.03       & -5.76$\pm$0.15        \\
HR 7631        & -13.49$\pm$4.37       & -26.13$\pm$0.81       & -13.8$\pm$2.4         & 1         & -13.81$\pm$0.5        & -27.64$\pm$0.09       & -14.41$\pm$0.28       \\
IS Eri         & -5.8$\pm$0.5          & -27.1$\pm$1.1         & -10$\pm$0.6           & 2         & -5.24$\pm$0.18        & -27.92$\pm$0.06       & -9.75$\pm$0.24        \\
LO Peg         & -5$\pm$0.3            & -23.8$\pm$0.9         & -15.7$\pm$0.7         & 2         & -6.57$\pm$0.43        & -29.07$\pm$1.58       & -13.15$\pm$0.61       \\
LP 745-70      & -7.5$\pm$0.6          & -27.9$\pm$2           & -12.6$\pm$0.6         & 2         & -6.73$\pm$0.47        & -27.85$\pm$0.07       & -12.95$\pm$0.21       \\
LTT 5826       & -10.86$\pm$2.79       & -19.21$\pm$2.86       & -10.85$\pm$0.34       & 1         & -10.3$\pm$0.1         & -18.52$\pm$0.1        & -12.62$\pm$0.02       \\
NLTT 19987     & -10.1$\pm$1.3         & -25.1$\pm$2.3         & -11.8$\pm$1.4         & 5         & -14$\pm$0.51          & -28.38$\pm$0.28       & -12.36$\pm$0.43       \\
PW And         & -5.42$\pm$0.33        & -28.69$\pm$0.63       & -17.94$\pm$0.74       & 1         & -4.43$\pm$0.22        & -27.75$\pm$0.46       & -15.68$\pm$0.31       \\
UY Pic A       & -7$\pm$0.2            & -28$\pm$0.2           & -14.5$\pm$0.2         & 2         & -7.27$\pm$0.03        & -28$\pm$0.12          & -14.76$\pm$0.08       \\
UY Pic Ba      & -10$\pm$1.3           & -25.6$\pm$1.2         & -14.9$\pm$1.3         & 2         & -7.48$\pm$0.25        & -27.08$\pm$0.89       & -13.38$\pm$0.57       \\
UY Pic Bb      & -10$\pm$1.3           & -25.6$\pm$1.2         & -14.9$\pm$1.3         & 2         & -7.48$\pm$0.25        & -27.08$\pm$0.89       & -13.38$\pm$0.57       \\
V0383 Lac      & -7.06$\pm$1.43        & -22.19$\pm$0.34       & -3.9$\pm$0.86         & 1         & -9.65$\pm$0.04        & -23.44$\pm$0.23       & -4.86$\pm$0.03        \\
V0577 Per A    & -5.8$\pm$0.5          & -26.5$\pm$0.8         & -15.9$\pm$0.6         & 2         & -6.53$\pm$0.17        & -27.84$\pm$0.11       & -16.57$\pm$0.04       \\
V0577 Per B    & -8.91$\pm$1.01        & -23.9$\pm$1.22        & -16.2$\pm$0.83        & 1         & -6.53$\pm$0.17        & -27.84$\pm$0.11       & -16.57$\pm$0.04 \\     
\hline

\end{longtable}
\footnotesize{$^{1}$\cite{montes2001}, $^{2}$\cite{Malo2013}, $^{3}$\cite{Maldonado2010}, $^{4}$\cite{Shkolnik2012}, $^{5}$\cite{Schlieder2012}}
\end{tiny}
\end{center}

\begin{tiny}

\begin{longtable}{lccccccccc}
\caption{Stellar parameters estimated of moving group members from the analysis with $\textit{Gaia}$ DR2 data.}
\\\hline  \hline
\multirow{1}{*}{Name}& \multirow{1}{*}{Moving Group} & \multirow{1}{*}{Prob} & \multirow{1}{*}{Sp. Type} & \multirow{1}{*}{G} & \multirow{1}{*}{Distance}& \multicolumn{4}{c}{Estimated parameters using $\textit{Gaia}$} \\
 & & & &(mag)& (pc) & Age (Myr) & $\it{U}$ (km~s$^{-1}$) & $\it{V}$ (km~s$^{-1}$) & $\it{W}$ (km~s$^{-1}$) \\ \hline \hline
\endfirsthead
\endhead
PW And & ABDor& 95 & K2 & 8.492 & 28.3$\pm$0 & 48.5$\pm$0.5 & -4.43$\pm$0.22 & -27.8$\pm$0.46 & -15.7$\pm$0.31 \\
2M 0019+4614 & ABDor & 63 & M8 & 16.802 & 38.3$\pm$0.2 & 95.5$\pm$0.5 & -- & -- & -- \\
2M 0032-4405 & ABDor & 66 & -- & 19.781 & 35$\pm$0.6 & 95.5$\pm$0.5 & -- & -- & -- \\
BD+61 119 & ABDor & 34 & F8 & 7.953 & 63.8$\pm$0.1 & 22.5$\pm$0.5 & -13.3$\pm$0.14 & -29.2$\pm$0.21 & -1.72$\pm$0.02 \\
\\
\hline
\end{longtable}
\footnotesize{Table 2 is published in its entirety in the machine-readable format. RA, DEC is also added in the machine-readable format.}
\end{tiny}

\begin{longtable}{lccl}
\caption{Comparison between the age of the moving groups obtained from this work with those listed in literature.}
\label{tab:my-table}\\
 \hline\hline
Name & Our estimate & Previous studies & Reference \\ 
&(Myr) &(Myr) & \\\hline\hline
\endfirsthead
\endhead
\multirow{2}{*}{$\epsilon$ cha} & \multirow{2}{*}{1.15-15.5} & 0.9-20 & \citet{Murphy2013}$^{a}$ \\
 &  & 8 & \citet{Torres2008}$^{b}$ \\ \hline
TW Hya & 1.75-27.5 & 10 $\pm$ 3 & \citet{Bell2015}$^{c}$ \\
 &  & 3-150 & \citet{Weinberger2013}$^{a}$ \\
 &  & $20^{+25}_{-7}$ & \citet{Mamajek2005}$^{d}$ \\ \hline
$\beta$ pic & 5.5-54.5 & 24 $\pm$ 3 & \citet{Bell2015}$^{c}$ \\
 &  & 26$\pm$3 & \citet{Malo2014}$^{e}$ \\
 &  & 10 & \citet{Torres2008}$^{b}$ \\ \hline
32 Ori & 6-17.5 & $ 22^{+4}_{-3}$ & \citet{Bell2015}$^{c}$ \\
\hline
Octans & 7.5-30.5 & 30-40 & \citet{Murphy2015}$^{e}$ \\
 &  & 20 & \citet{Torres2008}$^{b}$ \\ \hline 
Tuc-Hor & 3-60.5 & 45 $\pm$ 4 & \citet{Bell2015}$^{c}$ \\
 &  & 10-30 \& 40 & \citet{Kraus2014}$^{a,e}$ \\ \hline
Columba & 4-54.5 & $42^{+6}_{-4} $ & \citet{Bell2015}$^{c}$ \\
 &  & $\sim$40 & \citet{Torres2008}$^{b}$ \\ \hline
Carina & 3-34.5 & $45^{+11}_{-7}$ & \citet{Bell2015}$^{c}$ \\
 &  & $\sim$40 & \citet{Torres2008}$^{b}$ \\ \hline
Argus & 5.5-43.5 & 50$\pm$5 & \citet{Barrado2004}$^{e}$ \\ \hline
AB Dor & 7-54.5 & $149^{+51}_{-19}$ & \citet{Bell2015}$^{c}$ \\
 &  & 110 & \citet{Barenfeld2013}$^{d}$\\
 &  & 50-70 & \citet{Torres2008}$^{b}$ \\ \hline
Carina-near & 24.5-48.5 & 150-250 & \citet{Zuckerman2006} \\ \hline
Ursa Major & 5.5-48.5 & 414 $\pm$ 23 & \citet{Jones2015}$^{a}$ \\
 &  & 500 $\pm$ 100 & \citet{King2003}$^{a}$ \\ \hline
$\chi^{01}$ For & 5.5-27.5 & $~500$  &  \citet{Pohnl2010}$^{a}$ \\ \hline

\end{longtable}

\footnotesize{$^{a}$Based on isochrone fitting, $^{b}$Based on Li I 6707 \AA~EW and color-magnitude diagram}
{,$^{c}$Based on semi-empirical isochrone fitting,  $^{d}$Based on kinematical age estimation technique, $^{e}$Based on Lithium Depletion Boundary (LDB) technique.}

\newpage
\begin{longtable}{lccccc}
\caption{Moving Group characteristics obtained from our study.}\\
\hline
\hline
Group & \begin{tabular}[c]{@{}c@{}}Distance\\ (pc)\end{tabular} & \begin{tabular}[c]{@{}c@{}}Age\\ (Myr)\end{tabular} & \begin{tabular}[c]{@{}c@{}}U\\ (km$~s^{-1}$)\end{tabular} & \begin{tabular}[c]{@{}c@{}}V\\ (km$~s^{-1}$)\end{tabular} & \begin{tabular}[c]{@{}c@{}}W\\ (km$~s^{-1}$)\end{tabular} \\ \hline \hline
\endfirsthead
\endhead
32 Ori & 100 & 12.6 & -9.67 & -19.37 & -8.07 \\
ABDor & 36.4 & 30.34 & -7.63 & -27.35 & -12.86 \\
Argus & 29.5 & 21.96 & -19.98 & -14.97 & -8.62 \\
$\beta ~ pic$ & 33.7 & 19.38 & -10.65 & -15.88 & -9.15 \\
Carina & 78.1 & 16.64 & -9.42 & -22.5 & -3.94 \\
Car-near & 25.1 & 34.72 & -27.34 & -18.39 & -4.32 \\
$\chi^{01} For$ & 101.7 & 16.5 & -11.35 & -20.8 & -3.37 \\
Columba & 63.6 & 24.11 & -13 & -21.54 & -5.37 \\
$\epsilon$ ~ cha & 103.3 & 4.93 & -10.41 & -18.85 & -7.92 \\
Octans & 123.5 & 16.65 & -13.71 & -6.17 & -11.58 \\
Tuc-Hor & 46.2 & 19.75 & -10.72 & -20.2 & -0.7 \\
TW Hya & 62.7 & 6.53 & -14.64 & -18.1 & -5.85 \\
Uma & 25 & 26.47 & 14.02 & 1.83 & -10.04 \\
\hline
\hline
\end{longtable}

\end{document}